\documentclass[preprint,12pt]{elsarticle}
\usepackage{amsmath,amssymb,amsfonts}
\usepackage{algorithmic}
\usepackage{graphicx}
\usepackage{textcomp}
\usepackage{multirow}
\usepackage{array}
\usepackage{titlesec}  

\journal{Journal}

\begin{document}
\title{Ethics by Design: A Lifecycle Framework for Trustworthy AI in Medical Imaging From Transparent Data Governance to Clinically Validated Deployment}


 \author{Umer Sadiq Khan\corref{cor1}\fnref{label2}}
 \ead{umersadiq@hbeu.edu.cn}

 \affiliation[label2]{organization={Hubei Engineering University},
             city={Xiaogan},
          postcode={432000}, 
            country={China}}

\author{Saif Ur Rehman Khan*\corref{cor1}\fnref{label1}}
\ead{saifurrehman.khan@csu.edu.cn}

\affiliation[label1]{organization={Central South University},
             city={Changsha},
             postcode={932}, 
             country={China}}

\begin{frontmatter}
\begin{abstract}
\textbf{Context:}  
The integration of artificial intelligence (AI) in medical imaging raises crucial ethical concerns at every stage of its development, from data collection to deployment. Addressing these concerns is essential for ensuring that AI systems are developed and implemented in a manner that respects patient rights and promotes fairness.\\
\textbf{Objectives:}  
This study aims to explore the ethical implications of AI in medical imaging, focusing on five key stages: data collection, data processing, model training, model evaluation, and deployment. The goal is to evaluate how these stages adhere to fundamental ethical principles, including data privacy, fairness, transparency, accountability, and autonomy.\\
\textbf{Methods:}  
An analytical approach was employed to examine the ethical challenges associated with each stage of AI development. We reviewed existing literature, guidelines, and regulations concerning AI ethics in healthcare and identified critical ethical issues at each stage. The study outlines specific inquiries and principles for each phase of AI development.\\
\textbf{Results:}  
The findings highlight key ethical issues: ensuring patient consent and anonymization during data collection, addressing biases in model training, ensuring transparency and fairness during model evaluation, and the importance of continuous ethical assessments during deployment. The analysis also emphasizes the impact of accessibility issues on different stakeholders, including private, public, and third-party entities.\\
\textbf{Conclusion:}  
The study concludes that ethical considerations must be systematically integrated into each stage of AI development in medical imaging. By adhering to these ethical principles, AI systems can be made more robust, transparent, and aligned with patient care and data control. We propose tailored ethical inquiries and strategies to support the creation of ethically sound AI systems in medical imaging.
\end{abstract}

\end{frontmatter}
\section{Introduction}
\label{sec:introduction}
In the dynamic realm of artificial intelligence (AI) in medical imaging, ethics refers to the principal and guidelines that govern the responsible development and application of AI technologies. This development is growing in steps that are vital to ensure the patient well-being and uphold the high standards of the medical profession. It deals with concern; such as patient privacy, data protection, the accuracy and reliability of diagnostic tools, informed consent, and access to AI-augmented diagnostic services. Ethical considerations involve the assessment of transparency of AI systems', the integrity of AI algorithms, the accountability, and associated with decisions made using in medical imaging. The ethical criteria ensure the use of AI in a manner that upholds human rights and promotes the patient’s welfare.
The effectiveness of imaging data \cite{b1} has been improves as a result technological developments, Yet, this progress also lead to ethical concerns regarding their excessive misuse. The importance of obtaining informed permission cannot be underestimated, as it highlights the necessity to adequately convey the potential hazards, benefits, and alternatives related to imaging therapies. Professional ethics are the rules set for medical imaging \cite{b2} , important for understanding ethics, that govern how individuals in various disciplines intersect with one another and other involved parties. Therefore, to explore the ethical aspects of human nature, as well as the core goals of an organization, and gives a distinct portrayal of professionalism \cite{b3}. Big data \cite{b3,b4} and machine learning\cite{b5,b6} have drove in the development of imaging AI solutions that span from the entire value chain of medical imaging. The solutions of this type include image reconstruction \cite{b7,b8}, medical image segmentation \cite{b9}, image-based diagnosis \cite{b10}, and planning treatment \cite{b11}.  
It is expected that Artificial intelligence will play a crucial role in the development of medical imaging data \cite{b12}. The implementation of this technology can enhance medical imaging by augmenting the efficiency of collecting, processing, and interpreting medical images \cite{b13}. This can facilitate the extraction and incorporation of new information and imaging signature, leading to improved patient evaluation, prediction, and decision-making. Consequently, this helps healthcare \cite{b14} professionals diagnose and treat patients \cite{b15} with more efficiency and precision. 

However, despite significant improvements and developments in the medical imaging domain in recent years, the use and application of imaging AI technologies continue to be restricted in clinical practice. The digitization \cite{b16} of medical data and imaging has made data privacy issues of utmost importance. Medical physicians and professionals \cite{b17} are strongly encouraged to closely follow revised artificial standards that specifically deal with these changing ethical challenges, with the goal of encouraging openness and patient support in all imaging practices. The rapid growth of AI systems for medical applications \cite{b18} involves a careful balance. Evaluation models must possess high precision and certainty while also ensuring equitable for fluctuated patient populations. However, medical professionals and researchers have ethical rights the patients whose data is utilized for training the model. When sharing diagnostic models with unverifiable third parties, it is crucial to ensure that patient privacy is not violated \cite{b19}. 

Autonomous systems have become established as a database system that can scan medical images \cite{b13} using computers to track and loads the quantities of data. Back-propagation AI architectures have facilitated the application of medical imaging, leading to the possibility of correct diagnosis. Data ethics and medical images are closely intertwined \cite{b20}, as medical imaging relies on the dataset obtained from healthcare centers or research laboratories. The mechanism implemented \cite{b21} for collecting, processing, training, storing, and sharing data for medical imaging must prioritize the patient’s data privacy. 
The present use of AI in recent years has resulted in negative consequences at some point, including heightened infringement on privacy, biased decision-making based on data, and the absence of proper rationale for data-driven decisions \cite{b22}. AI is being applied in various domains, including virtual assistants, autonomous vehicles, translation software, virtual assistants, translation software, and the medical area, where it may resulting in considerable changes \cite{b16}.

Protection of patient data is a key issue in medical image analysis, which may include sensitive information like medical records and personal identification. To protect patient privacy \cite{b23} and maintain confidence in the healthcare system, it is necessary to assure security and protection of patient information \cite{b24} from unauthorised access. This entails refraining any bias in the in the admission of training data \cite{b25}, and ensuring that the analysis is not discriminating against certain groups or individuals. Therefore, the study should prioritize gathering and assessing of data ethics in medical image analysis to develop a few guideline for  researchers and practitioners to the systematic approach in ethical data collecting and analytics \cite{b6}.Though AI is getting attention in health care, literature on is missing through investigation into the specific security and privacy threats that are linked to medical data.

Our study seeks to offer ethics insights into various categories in AI system development for medical imaging data, such as data collection, data processing, model training, model evaluation, and deployment. This study also endeavours to fill the void by identifying security and privacy concerns and mapping them to the life process of AI systems in health care, from data collection via model training to clinical deployment. Through identification and then assessment these threats, we can gain insight into the vulnerability and related risk with the use of AI in health care. We aim to contribute to both theory and practice by highlighting the importance of tackling with these threats and offering mitigation measures. The findings of this study can shape the AI systems development in health care and contribute to the establishment of human-machine synergy and offer to health care organizations to comprehend the potential benefits and risks of using medical imaging in AI. This study explores the ethical implication of AI in medical imaging data throughout different stages, from data collection, data processing, model training, model evaluation, and deployment. It aims to scrutinise meticulously the current existing protocols and posit in to identify potential challenges and future threat on the horizon.

This article outlines ethical framework by proposing principles that should guide the development and implementation of the AI algorithms that can learn from and assessing data obtained in the routine provision of medical services. The study’s value resides in its capacity to enlighten policy makers, health care organizations, and AI developers about the security and privacy challenges faced while medical imaging data in different phases is being executed. The findings of this study offer ethical recommendations to ensure the safety and ethical use of AI in health care. With careful governance, the benefits of AI models can be realized while safeguarding patient data and public trust. Ultimately, this study contributes to the advancement of knowledge in the field of AI in health care and supports the development of secure and privacy-preserving AI medical imaging data.

\subsection{Current Ethical Practices, Challenges and Limitations}

The dynamic growth of artificial intelligence in medical imaging presents a future filled with immense possibilities \cite{b26}. As AI becomes more integrated into healthcare systems \cite{b27}, its applications in medical imaging expand dramatically. However, this rapid development also introduces significant ethical challenges.  

The concept of ethics \cite{b28}, a branch of moral philosophy, examines and seeks to elucidate issues concerning virtue, good and evil, and free will. Ethics itself evolves alongside human civilization and is not universally consistent across different cultures and social structures. \cite{b29} A comprehensive understanding of the role of ethics and its impact on regulatory standards in artificial intelligence (AI) for medical imaging necessitates a reliable interpretation of humanitarian concepts from moral philosophy within the context of modern technologies that support AI and its applications in medical diagnostics and treatment planning. The traditional principles of medical ethics like patient confidentiality, informed consent \cite{b23}, and non-malfeasance are tested as AI reshapes diagnostic and treatment processes. As we advance into this new era, it is crucial to tackle these ethical issues to fully leverage AI in medical imaging while upholding the fundamental values that underpin patient care and medical practice.

Ethical considerations \cite{b30} in artificial intelligence for medical imaging are closely connected with data ethics \cite{b3}, as the effectiveness of AI technologies depends on the quality and integrity of data derived from healthcare providers and research facilities. It is important to assure that the approaches of data collection \cite{b3,b6}, storage, and dissemination are in compliance with patient privacy. At present no universally accepted or standardized ethical framework for data practices in this domain. A major concern in AI-driven medical imaging is privacy concern which covers details like personal identifiers and medical histories \cite{b31}. Security measures should be robust enough within the health care system to prevent unauthorized access, maintain patient privacy, and uphold trust. Moreover, the selection of data used to train AI models should be unbiased to avoid discriminatory analyses.

The collecting data for AI in medical imaging is a complicated yet essential process \cite{b32}. The principal of dealing with data, shared, and sometimes reused is an important feature to modern medical research that is being more regularly motivated by advancements in AI and data science \cite{b33}. Although, present debates regarding the ethics of medical data manipulation in several contexts, there is still a significant void of support in addressing with ethical concerns that this challenging area poses. Privacy and anonymity are of paramount importance to ensure the ethical data collection of data for AI applications in medical imaging \cite{b5,b6}. It ensures privacy and anonymity of patients involved. This entails informed consent for the collection of their medical data and privacy and keep secure their personal information. The technique employed for data collection and medical data analysis must be accurate, reliable, and universally relevant to the demographics they aim to serve. AI is being used in various domains, such as autonomous vehicles, virtual assistants, translation software, and the medical area, where it has proved its advantages. The current ethical practices of AI for medical imaging emphasize ensuring patient privacy and data security. While data collection, ethical protocols include anonymizing patient data to mask the identities, therefore making sensitive information cannot be linked back to individuals. Data security in data processing, involves securing data, unauthorized access, and breaches is paramount, using encryption and secure data storage solutions.

During the phase of training the model, it is essential to adhere to ethical principles by implementing dataset that are both balanced and diverse \cite{b34}. This approach helps to minimize the potential for bias in the trained model. In the domain of machine learning \cite{b35}, algorithms function by utilizing various datasets, often referred to as training data, to assess the correct outputs for individuals or entities. In addition, training data form the key basis upon which the algorithm develops its model, allowing it to predict appropriate outcomes or entities with respect to ethical considerations including fairness, transparency, and privacy \cite{b36}. For model evaluation \cite{b37}, thorough testing on different datasets ensures the model is performing well with different demographics. Finally, deployment practices embrace continuous monitoring promises the model reliability and adherence to medical standards.

\subsection{Challenges in Medical Imaging}
The development of AI system for medical imaging poses several ethical challenges. Obtaining informed consent can be significant challenge during the data collection process, especially when dealing sensitive data or data from minors or vulnerable populations. Preserving anonymity is an essential concern when navigating deep and intricate dataset through data processing. Additionally, training dataset frequently possess biases, which could end up in the development of algorithm that showcase suboptimal performance underrepresented groups. The assessment of models also poses ethical challenges, ensuring fairness and erasing bias is complex because of the lack of standardized benchmark and varying testing conditions. Ultimately, during the deployment phase, the reliance on AI necessitates substantial human oversight. AI algorithm can give rise to ethical challenges when their results contradict with clinical findings.

\subsection{Ethics Limitations}
The ethical limitations of AI in medical stem from several factors. These factors give rise to ethical limitations, such as limited access to extensive, external datasets, which in turn limits the practicality of AI models in medical imaging. The limitation is exacerbated by privacy concerns, which hinder the sharing of medical data among various groups. The depiction of varied groups is frequently insufficient, resulting in AI models that may not exhibit equitable performance across multiple demographic grouping. Ensuring patient confidentiality and preventing illegal access to sensitivity health information while handling AI. The anonymity of AI decision making process of possesses difficulties for user and regulators in comprehending and placing trust in the outputs generated by AI. Improper use of AI might pose risk to patients, perhaps resulting inaccurate diagnoses and unsuitable treatment suggestions. The possibility of this misuse highlights the necessity for strict regulatory supervision and ethical principles in the advancement of AI in medical imaging.

\section{Ethics for AI in the Development Phase}
AI can provide broader access to medical imaging data knowledge. However, despite recent improvements in factual accuracy the recurring issue of ethical information in the AI development such as ethical consideration, data privacy \& integrity, fairness, unbiased algorithm, accuracy and reliability, and monitoring \& update are potentially harmful consequences remains. For this, we propped an ethical model as shown in Fig. 1, during the AI development for medical imaging data. However, it’s not the same with the conventional, currently available studies as it highlights some of the important ethical aspects concerned with AI development for imaging data. Moreover, to the best of our knowledge, there is no one study that highlights the collective problems of AI development system at each step with respect to ethical considerations. In addition, all of the review studies related to ethics in AI development are in constant flux; therefore, we cannot adopt the conventional selection criteria for selecting the studies. Moreover, every execution of the proposed model encounters a referral and many more pertinent ones are included in specific areas of interest, for instance the issues addressed. 

\begin{figure*}[!t]
\centerline{\includegraphics[width=1.2\textwidth]{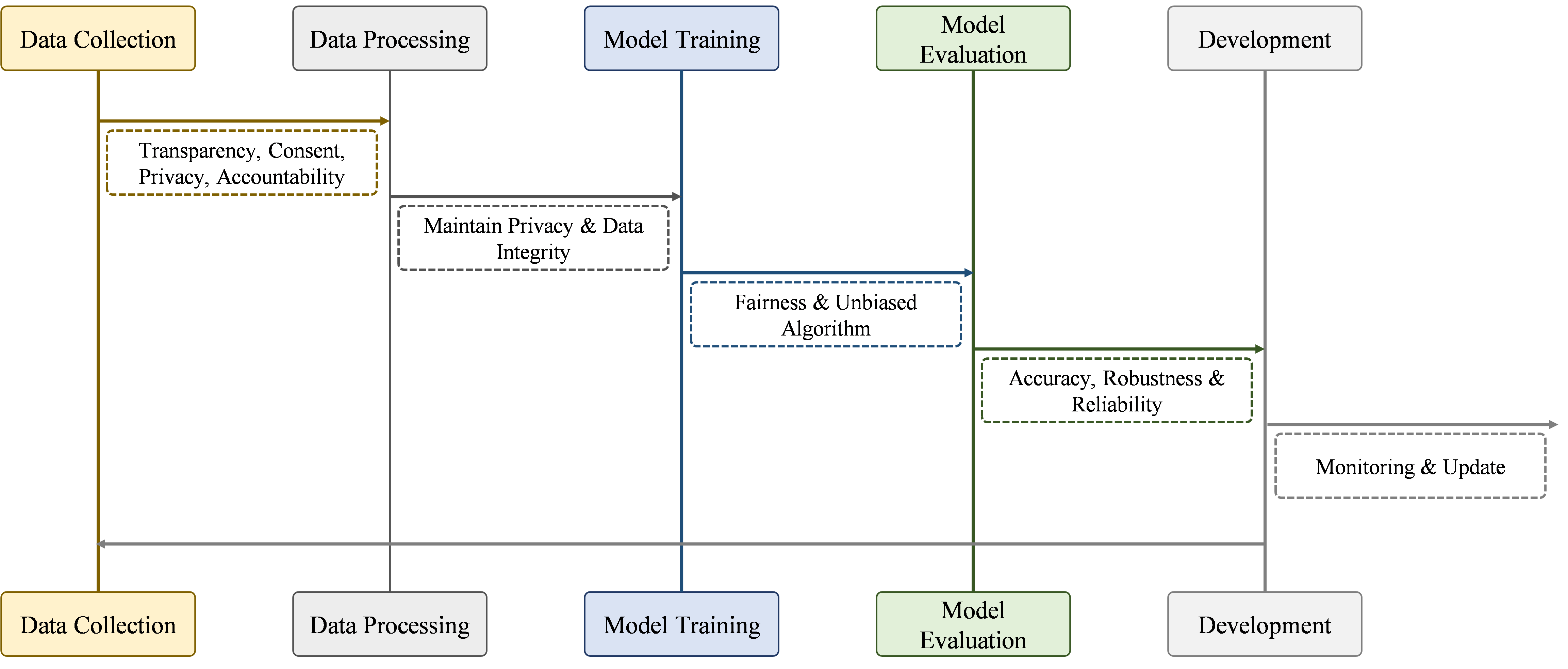}}
\caption{Visual encapsulation analysis dissects ethics through the lens of five pivotal stages: data collection, data processing, model training, model evaluation, and deployment, emphasizing adherence to key ethical principles.}
\label{fig1}
\end{figure*}

\subsection{Data Collection}
It is imperative from an ethical standpoint to ensure that all patients provide informed consent for the use of their medical images \cite{b38}. This includes clearly explaining the intended use of the images, potential risks involved, and affirming the patients' rights to withdraw consent at any time. One should implement measures to protect patient privacy \cite{b39}, including de-identifying images to remove any personal identifiers and securing data storage and transmission systems. Data collection must be conducted in an ethical and transparent manner, ensuring adequate representation of protected groups \cite{b40}. Implement encryption and secure networks to protect data from being unauthorized access, while consistently ensuring that is kept up-to-date security measures and performing security check as needed. Comply strictly with the applicable laws and regulations regarding the use of medical imaging such as the HIPAA (health insurance portability and accountability act the U.S law) \cite{b41,b42} or GDPR (general data protection regulation in Europe) \cite{b43} to ensure that the data collected are in compliance with the laws. It is legal and appropriate to seek the permission of an ethics committee or institutional review board (IRB) before embarking data collection so as to uphold the rights of the patients. To maintain the quality of research, one should limit the collection of data to what are relevant, specific research objectives of the study, but not more than that. Inform all stockholder on how data is collected, used, stored, and with whom is shared. Protect the rights of all the participants and ensure data collection processes does not harm the participants. Following these steps will assist in the creation of ethical guidelines for the collection of medical imaging data that respect right of the participants as well as legal and ethical framework. 

\subsection{Data Processing}
The paramount goal is to maintain and ensure transparency in the transformation and careful pre-processing of data before it is fed into the processing. The ethics highlight the responsibility make sure all personal identifiers are stripped off from the data to maintain patient privacy \cite{b44}. Employing techniques such as anonymity where appreciate to enable research on the information and maintaining confidentiality. Limit the admission of sensitive information to authorized personnel only \cite{b6}, implementing role-based access controls and regularly reviewing access permissions to ensure they are appropriate. Preserve the accuracy and completeness of medical images during processing, incorporation of control to prevent data manipulation from leading to errors or change the clinical relevance of the images. Employ encrypted storage and transfer technique to prevent unauthorized access to data, ensuring that any third-party services may be used for data storage or processing also meet with relevant security compliance. Adhere to international standards and recommendation for medical image processing, for instance the DICOM(Digital Imaging and Communications in Medicine)\cite{b45} Standards Committee and ISO International Organization for Standardization \cite{b46}. Keep records of all activities, involving data processing and the parties that engaged in the activity. Regularly review data processing activities through an ethics committee or similar body to ensure on-going compliance with ethical and legal standards. Be transparent about data processing practices and provide reports to stakeholders, including participants, if requested, to promote trust and accountability. Ensure proper plan are in place in case of data breaches such as notifying the affected individuals and regulatory bodies where required. One the same note, data is disposed of more securely and in a manner that is complaint with the law. These steps assist in preventing the compromised of ethical benchmark in the processing of medical imaging data \cite{b47}, patient rights should be protected and the data is used appropriately and securely.

\subsection{Model Training}
Training AI models for medical imaging requires a high level of care and sensitivity with rules that govern patient rights care, data security, and the overall integrity of the medical process \cite{b48}. Commencing using a dataset is important for the following reason that is also high-quality and relevant but also contain data needed for analysis. This make sure that results of the model are reliable and applicable across various demographics, reducing health disparities. Informed consent is paramount. Participants must possess comprehensive knowledge on the utilization of their data, especially in contexts they may not anticipate, such as AI training \cite{b42}. Make sure that all use cases are covered in the consent forms respects patient autonomy and legal requirements. Privacy protection cannot be overstated \cite{b49}. By using latest  anonymization and pseudonymization tools to ensure data cannot be traced back \cite{b50} to individuals is essential for maintaining trust and complying with legal frameworks like HIPAA or GDPR. The AI model should be transparent on its own part. Detailed information on how the model processes data, makes decisions, and has been checked for bug or biases should be easily accessible. This transparency is important \cite{b54} for healthcare providers and patients who use the model's results for decision making. Ethical oversight should not cease as soon as the model has been established. There is always the need to continuous monitoring and updating is necessary to adapt to new data, emerging biases, or changes in clinical principals. Regular scrutiny of an ethics committee consisting a diverse group of stakeholders from ethicists to patient advocates ensures the model remains a reliable model in clinical practices. Lastly, the thereby any potential impacts of deploying such AI models must be well weighted. Conducting the detailed analysis to determine the impact of the model on the care of patient can ensures that the technology is applied appropriately and is indeed beneficial to healthcare provision.

In conclusion, the best practices in ethical AI training in medical imaging encompasses strict procedures and around rigorous processes and continual supervision to ensure patient interests, ensure data privacy, and deliver dependable, unbiased healthcare solutions.

\subsection{Model Evaluation}
In an ethical Perspective, it is necessary to assess the predictions of the model in a manner that is clear and trustworthy to other healthcare professionals and other stakeholders who are interested in accessing such information. Metrics are in concordance with patient health outcomes , and safety considerations \cite{b52,b53}. In this stage, to retain accuracy and reliability from an ethical standpoint involves rigorously testing the AI models to ensure how well they work in all relevant scenarios and patient groups. From an ethical perspective, it is vital to validate the model's effectiveness thoroughly to hinder potential harm that may be associated with inaccurate or unreliable forecast information. This includes performing thorough testing on diverse datasets that are realistic to identify and address any imbalance in performance. Ethical considerations also require transparency and specific reporting the model's shortcoming and performance metrics, it ensuring that stakeholders are fully are aware of the strength and limitations of the AI system. Furthermore, on-going monitoring and re-evaluation of the model post-deployment are crucial to update its accuracy and reliability with the newly data becomes available and clinical practices evolve. This commitment to high standards in model evaluation protects patient interest and trust placed on in AI-driven medical technologies.

\subsection{Deployment}The deployment step of monitoring and updating AI models in medical imaging entails substantial ethical responsibilities. From an ethically, it is crucial to emphasize that monitoring should be essential to ensure that the AI system function properly and remains robust and secure in a real-world clinical environment \cite{b54}. This entails constantly checking the model's effectiveness and its contribution impact on patient outcomes \cite{b55}, and being alert for any trends of decline in accuracy or emerging biases when the model continues to deal new patient data. Updating the AI model is also crucial aspect that need to be considered while designing the system \cite{b56} responsibility to evolve to new knowledge in the field of medical, patient demographics, and clinical practices. It helps keep the technology stays current and keeps advancing to meet the high standards of medical care. Some of the concern include: lack of transparency on how the AI model is monitored and updated is crucial for maintaining trust \cite{b57} among healthcare providers, patients, and regulatory bodies. Moreover, it is essential to ensure communication about any changes made to the AI system, the causes behind this alteration, and their potential consequences on performance and patient. Finally, including a diverse group of stakeholders, including clinicians, ethicists, and patients, in the monitoring and updating process make sure that the AI system general and societal professionalism and ethical principles, further enhancing its acceptance and integration into healthcare.

\section{Considerations, Accessibility Levels, Interrelated integrated Ethics, and Ethical Quality Check for Future}

In the context of medical imaging, the development of AI entails a comprehensive progression commencement with data collection. Ethical questions are the highest priority within this stage. Data is collected with the consent and it should be expressed in various forms. The next step is data processing where the data is cleaned and formatted simultaneously preserving privacy and avoiding biases. Model training ensues; the AI algorithms are trained from the processed data under strict ethics oversight to ensure that there is no biased result. Model evaluation is an important phase to determine the fairness and accuracy of the AI, prior to the deployment to ensure they meet ethical and clinical standards. Finally, the deployment phase that integrate clinical settings of AI and its effectiveness is tracked constantly to ensure it complies to ethical norms and improve patients’ outcomes without intruding into their privacy or autonomy. In each stage, the process is conducted based on the moral values that make the AI systems not only effective, reliable, and also respect to patient rights. 

\subsection{Ethics Considerations}
The factors outlined in Fig. 2 highlight the importance of placing ethical principles into practice at each phase of AI development in the medical imaging domain. The figure shows breakdown of each step, starting from the data collection to deployment includes specific ethical considerations such as consent, data security, model transparency and fairness. In addition, it reveals which stages are accessible for public/private and third-party entities. This structured approach ensure that the development of medical imaging technologies is in line with high ethical standards, thus, preserving the integrity and trust necessary for sensitive applications like healthcare.

\begin{figure*}[!t]
\centerline{\includegraphics[width=1.2\textwidth]{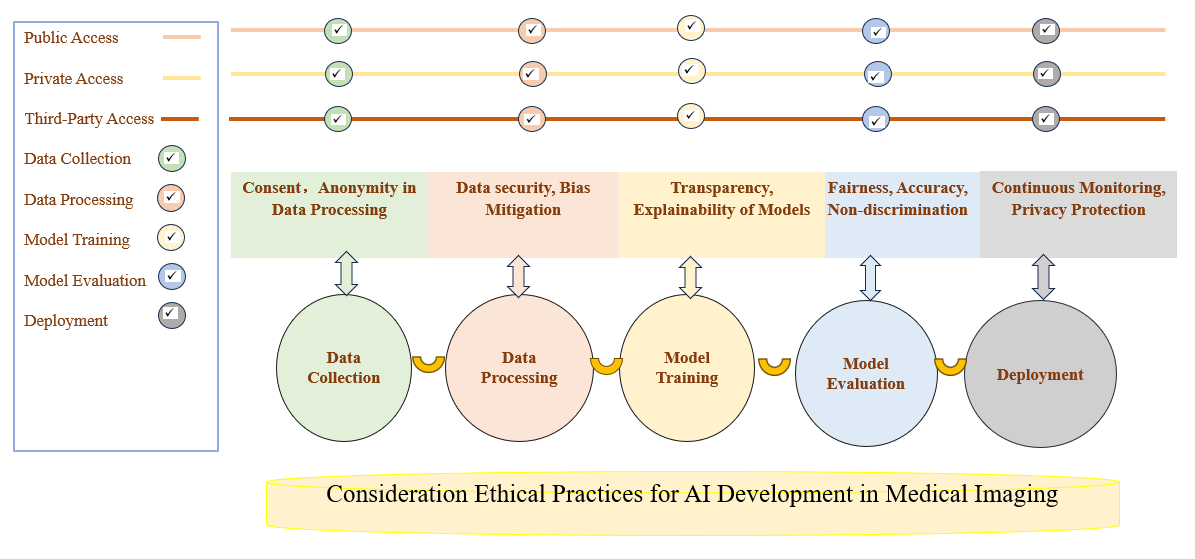}}
\caption{Ethics considerations for public access, private, and third-party access in AI development for medical imaging.}
\label{fig2}
\end{figure*}

The major ethical concern that arises during collection phase, include informed consent, keeping patient data anonymity, and sample diversity. Data can be sourced from public, private or a third party; each of these sources has its own strict adherence to ethical guidelines that have to be followed to protect the rights and patient privacy. For processing data, appreciate measure need to secure the data and deal any inherent biases that may lead to biased AI model results. Ethical norms are followed regardless of the nature of the data's source. It is thus important to make AI training processes procedures and decision-making process more transparent so is to gain trust in the development of AI systems. Thus, opens up the possibility for external scrutiny and validate by different stakeholders like public researchers and private auditors. Ethics issues in model evaluation are primarily concerned with the bias and accuracy of AI models, avoiding discrimination across different demographics through rigorous testing. Post-deployment constant monitoring of AI systems to make sure that they follow ethics and adjusted without violating the privacy or ethical norms.

\subsection{Ethics Accessibility Level}
Table I on the accessibility levels for public, private entities, and third parties in AI development stages in medical imaging is created to delineate who can access critical data ethics at various stages of the AI development lifecycle. This is important for several reasons:

\begin{table}[h!]
\centering
\caption{Ethics accessibility level for data in AI development stages in medical imaging.}
\label{table}
\resizebox{\textwidth}{!}{%
\begin{tabular}{p{3.5cm}p{2.5cm}p{4.2cm}p{1.5cm}p{1.5cm}p{2cm}}
\hline
\textbf{Stages of AI Development} & \textbf{Ethically} & \textbf{Description} & \textbf{Public} & \textbf{Private} & \textbf{Third Parties} \\ \hline
Data Collection & Restricted & Access to medical imaging data is ethically limited due to privacy concerns and regulatory requirements. & No & No & Limited \\ \hline
Data Processing & Limited & Data may be more ethically accessible but still tied to specific research initiatives or institutions. & Yes & No & Limited \\ \hline
Model Training & Moderately Open & Ethical collaborations are encouraged to access data with necessary partnerships and agreements in place. & Yes & Limited & Yes \\ \hline
Model Evaluation & Open & Data for ethical validation might be more openly accessible, often in search of validation partners to ensure thorough assessments. & Yes & Yes & Yes \\ \hline
Deployment & Restricted & Access to deployment-level data is ethically reserved to ensure patient privacy and compliance with medical regulations. & No & No & Limited \\ \hline
\end{tabular}%
}
\end{table}

Table I serves to integrate ethical considerations directly into the management and access of data across various stages of AI development. Here’s a breakdown of its importance and functionality.

\subsection{Ethical Foundations}
Table integrates ethical accessibility into the framework, make-sure that all decisions about data usage are ethical and align with core ethical principles, such as patient autonomy respect, confidentiality is maintained, and fair access of data.

\subsection{Stages of AI Development}
Data collection has the highest level of restriction, preserve patient privacy from the outset, thus ensuring that data collection follows ethical standards and regulatory requirements. The fact that the third parties are only allowed to have a limited access is another sign of an attitude of caution towards the handling of sensitive data. The limited access for third parties further emphasizes a cautious approach towards the handling of sensitive data. Even though the data processing step makes the data a bit more accessible, it is still limited and strictly controlled. This ensures that the data processing, which may include aggregation, anonymization or analysis, is not through compromise of ethical standards. The model training is moderately open; thus, it facilitates the innovation and the development of medical AI technologies through ethical collaborations. Partnerships and agreements are the key here to specify the clear limits and responsibilities, thus ensuring that all parties adhere to agreed ethical practices. Validation is transparent and can involve multiple stakeholders to ensure that the AI models perform as expected and restricting biased or adverse effects. Open validation is an essential component of transparency and fairness, because it allows various inputs to evaluate the model’s effectiveness and bias. Similar like data collection, deployment is also bounded to ensure that when AI systems are employed in real-world settings, they do not infringe on patient privacy or exceed the ethical norms set during development.

\subsection{Purpose of Ethical Guidelines}
Protection of privacy
\begin{itemize}
	\item[$\bullet$] Make sure that AI process has to be designed to safeguard the personal and data sensitivity from leakage at each step of the process.
  \end{itemize}
  
Ensuring accountability
\begin{itemize}
	\item[$\bullet$]Specifying the rules that have access to what data at each stage, the table provided accountability to all the involved parties.
   \end{itemize}
   
Promotion of transparency
\begin{itemize}
	\item[$\bullet$]Stages as validation that are open assist in foster trust among the public and stakeholders, since the process and outcomes of the AI models are known.
   \end{itemize}
   
Supporting collaboration 
\begin{itemize}
	\item[$\bullet$]The principal of the ethics and accessibility promotes safe and collaboration constructive, this makes innovations in medical imaging both technologically and ethically appropriate.
   \end{itemize}
   
By applying these ethical standards in the various phases of AI development, make sure organizations practices are not only legal but also uphold the highest ethical standards; thereby maintaining both the integrity of their work and the trust of the public and other stakeholders.

\section{Interrelated Integrated Ethics}
Fig. 3 outlines how ethical considerations are integrated throughout the stages of AI development in medical imaging. Ensuring the technology not only advances healthcare but also upholds the highest standards of ethical practice. These measures are essential to maintain trust among users, patients, and the broader community. The measures also ensure that AI applications in healthcare are safe, fair, and respectful of privacy and autonomy.

\begin{figure}[!t]
\centerline{\includegraphics[width=0.8\textwidth]{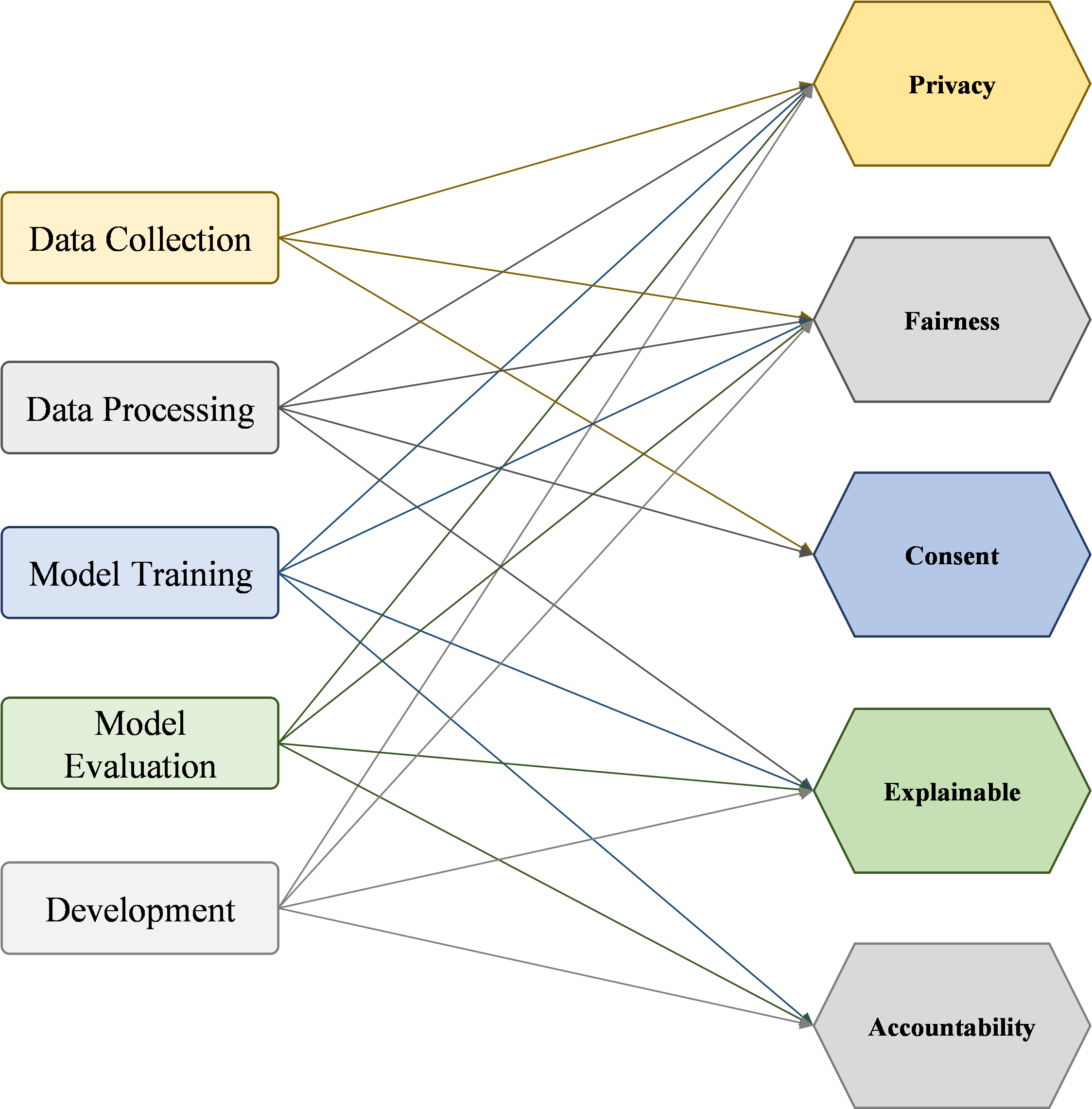}}
\caption{ Interrelated integrated ethics across the AI development in medical imaging.
}
\label{fig3}
\end{figure}
Explanations of each stage in the interrelated integrated diagram are given below;

\subsection{Data Collection}
\begin{itemize}
    \item[$\bullet$]Guaranteed by the anonymization and secure data handling.
    \item[$\bullet$]Attempts are made to create different data sets.
    \item[$\bullet$]Patients are informed and obtained their consent for the use of data.
   \end{itemize}

\subsection{Data Processing}
\begin{itemize}
    \item[$\bullet$]It entails on-going of robust security measures.
    \item[$\bullet$]Data is dealt to ensure it still reflects representative of diverse populations.
    \item[$\bullet$]Ensure algorithms are transparent and explainable as much as possible.
    \item[$\bullet$]Continued strict adherence to initial consent regarding data usage.
   \end{itemize}

   \subsection{Model Training}
\begin{itemize}
    \item[$\bullet$]Persisted emphasis on secure data management.
    \item[$\bullet$]Constant adjustments to the development models to continuously minimize bias and improve fairness.
    \item[$\bullet$]Emphasis model decisions comprehensible by external parties.
    \item[$\bullet$]Make mechanism to accountable developers and users for model performance and impacts.
   \end{itemize}

     \subsection{Model Evaluation}
\begin{itemize}
    \item[$\bullet$]Data used in the evaluation should be kept valid and secure.
    \item[$\bullet$]Processes of evaluation are designed to find and fix the bias in models.
    \item[$\bullet$]Clear specification of the evaluation criteria and procedures.
    \item[$\bullet$]Before deployment, robust processes to review and audit model performance.
   \end{itemize}

    \subsection{Deployment}
\begin{itemize}
    \item[$\bullet$]Utmost level of data protection is guaranteed as models are applied in the real clinical application.
    \item[$\bullet$]Control and tune-up to make the real-life application fair forever.
    \item[$\bullet$]Systems put in place to ensure on-going oversight and responsibility for deployed models.
   \end{itemize}

\subsection{Rationale for Non-applicability Ethics during Interrelated Integrated Ethics }
In the process of creating interrelated Integration Ethics throughout the AI development for medical imaging, some stages may not be applicable because of a special context or regulatory requirements. Here we explore why these stages are not applicable in certain scenarios.

\subsubsection{Transparency \& Explainability}

Data Collection and Deployment: At these stages, the main concern is usually on privacy and security rather than on the transparency of processes or algorithmic decision-making. In data collection, the main concern is how to gather and secure the data while in deployment; the primary focus might turn towards ensuring the model operates effectively and safely within its intended environment while in deployment. Transparency and explainability are more of a hindrance at these stages rather than in model training and evaluation, it is crucial where the understanding how models arrive at decisions.

\subsubsection{Accountability}

Data Collection and Data Processing: The accountability mainly revolves around to the observance of data protection laws and making sure that data handling processes are compliant and secure. The wider implication of the model’s behaviours and outcomes, more relevant considered to accountability, focus of intension only during the later stages of model training and evaluation, when its decisions and their impacts are more pronounced and assessable.

\subsubsection{Consent \& Autonomy}

Model Training, Model Evaluation, and Deployment: At these stages, initial consents that were obtained during the data collection phase are presumed to cover these activities. Hence, the attention on obtaining additional consent might be limited emphasized unless new usage scenarios or significantly different data handling processes are introduced. In model training and evaluation, probably more emphasis will shift to internal review processes and ensuring that the models behave as intended without having to steps further consent, assuming that the use remains within the bounds of the original consent terms. 

The design of the Fig. 3 showcases a strategic effort to addressing the most pertinent ethical concerns at every phase; thus, make sure those resources and efforts are intensive where they are most needed and highly effective. This proposed approach helps in managing the practical facets of practicing ethical guidelines with ease without diluting avoid on too many areas simultaneously.

\section{AI Ethical Quality Check for Future}

Table II depicts a set of ethical probes specifically built-in for every phase of the development cycle of AI in medical imaging. These levels cover data collection, data processing, model training, model evaluation, and deployment. The sole aim of the questions to ensure that ethical considerations are systematically integrated and addressed into each step, promoting responsible and trustworthy AI system development.
\begin{table}[!htbp]
\centering
\caption{Ethical quality checks for future; tailored to each stage at the AI development cycle in medical imaging.}
\label{table}
\begin{tabular}{p{3cm}p{3.5cm}p{7.5cm}}
\hline
\textbf{Stage of Development} & \textbf{Ethical Aspect} & \textbf{Ethical Question} \\ \hline
\multirow{2}{*}{Data Collection} & Fairness & Have diverse demographic groups been represented in the dataset? \\ \cline{2-3}
& Data Privacy \& Security & How is patient confidentiality protected during data collection? \\ \hline
\multirow{2}{*}{Data Processing} & Transparency \& Explainability & How are the data processing steps documented for transparency? \\ \cline{2-3}
& Bias and Fairness & What methods are in place to detect and mitigate bias during data processing? \\ \hline
\multirow{2}{*}{Model Training} & Accountability & How are the decisions made by the model during training tracked and reviewed? \\ \cline{2-3}
& Informed Consent & Are the sources of training data clearly disclosed and is consent obtained where necessary? \\ \hline
\multirow{2}{*}{Model Evaluation} & Transparency \& Explainability & Is the model's performance evaluated in a way that is understandable to non-technical stakeholders? \\ \cline{2-3}
& Clinical Relevance & Are the evaluation metrics aligned with clinical outcomes and patient safety? \\ \hline
\multirow{3}{*}{Deployment} & Continuous Monitoring & Is there a plan for continuous monitoring of the AI system in a real-world clinical setting? \\ \cline{2-3}
& Regulatory Compliance & Does the deployment plan include adherence to relevant healthcare regulations and ethical standards? \\ \cline{2-3}
& Risk Assessment & Prior to deployment, how are risks assessed and communicated to all stakeholders involved? \\ \hline
\end{tabular}
\end{table}
All the question is designed to prompt developers and stakeholders to contemplate the ethical that is, the specific dimensions, at the particular development stage of the AI system, so that final product is ethically sustainable and aligned with the best practices of patient care and data stewardship.

\section{Application of Software in Engineering: AI in Medical Imaging}
Software, especially when combined with artificial intelligence (AI), plays a crucial role in modern engineering fields, with profound impacts in healthcare and medical imaging. Engineering disciplines, including software engineering, electrical engineering, and biomedical engineering, utilize AI-driven software to enhance the accuracy, efficiency, and effectiveness of systems and applications.

In the field of biomedical engineering, software integrates with medical imaging technologies to process vast amounts of imaging data, enabling doctors to diagnose and treat patients more accurately. For example, AI-based software tools are employed in radiology to analyze medical images such as X-rays, CT scans, and MRIs. These tools assist in detecting anomalies like tumors, fractures, or diseases that may otherwise go unnoticed by the human eye. The algorithms are trained on large datasets, leveraging machine learning models to learn patterns and make predictions, all of which significantly reduce the time required for diagnostic procedures.

Moreover, software in medical imaging systems has also contributed to personalized medicine, where AI algorithms are used to tailor treatment plans based on individual patients' diagnostic data. For example, AI-driven image segmentation tools allow radiologists to identify and analyze specific areas within an image, helping to assess the severity and progression of a condition. This technology also enhances the ability to monitor patients over time, offering insights into the effectiveness of treatment options and adjusting care plans accordingly.

Beyond data processing, the software also plays an integral role in model evaluation and deployment. In medical imaging, the evaluation process involves validating AI algorithms against real-world clinical data to ensure accuracy and reliability before being deployed in hospitals or clinics. Transparency and explainability are crucial at this stage, as clinicians must trust the system's decision-making process. Ongoing deployment assessments ensure that AI systems continue to comply with ethical standards and maintain their effectiveness throughout their lifecycle.

\section{Conclusion}
In this detailed study, we explore the ethical principles essential for the application of artificial intelligence (AI) in medical imaging, covering every critical phase. Our analysis dissects ethical considerations through the lens of five pivotal stages. It covers data collection, data processing, model training, model evaluation, and deployment while focussing on the key ethical principles including data privacy, eliminating bias, transparency, accountability, and patient autonomy. The paper elaborates ethical consideration such as securing patient consent and anonymizing personal data to safeguard privacy at the data collection phase. Our analysis has shown that ethical considerations are essential at every phase of the process. This entails proper ethical data collection techniques. Firstly, it involves maintaining privacy and data integrity while processing the data. Second, it encourages fairness and non-biased models during training.  Third, it focuses on the accuracy and reliability of the model evaluation, and executing robust monitoring and updating protocols during its deployment phase. Additionally, we propose a tailored set of inquiries for each stage, designed to promote the ethical advancement of AI in medical imaging, while robust alignment with superior standards of patient care and data governance. This study set a comprehensive ethical framework for AI deployment in healthcare, aiming to inspire future developments would adhere to ethical principals in medical technology innovation.\\\\
\textbf{Ethics approval and consent to participate}\\
Not applicable\\\\
\textbf{Consent statement}\\ 
Not applicable\\\\
\textbf{Conflict of Interest} \\
All authors of this study declare that they have no conflict of interest. \\\\
\textbf{Data Availability}\\
The data that support the findings of this study are available from upon reasonable request and with permission of the author.\\\\
\textbf{Acknowledgement}\\
This work was supported in part by the National Natural Science Foundation of China under Grant 61972136, in part by the Natural Science Foundation of Hubei Province under Grant 2020CFB497, in part by the Hubei Provincial Department of Education Outstanding Youth Scientific Innovation Team Support Foundation under Grant T201410 and Grant T2020017, in part by the Natural Science Foundation of Education Department of Hubei Province under Grant B2020149, in part by the Science and Technology Research Project of the Education Department of Hubei Province under Grant Q20222704, and in part by the Natural Science Foundation of Xiaogan City under Grant XGKJ2022010095 and Grant XGKJ202201 0094.\\\\
\textbf{Funding}\\
No Funding\\\\
\textbf{Authors contribution}\\
Umer Sadiq Khan \& Saif Ur Rehman Khan conceived the methodology design, design of the study, data collection, implemented the experiments, analysis and interpretation of the data, and drafting of the manuscript. Contributed to data collection, analysis and interpretation of the data, and critically reviewed the manuscript. 
All authors have read and agreed to the published version of the manuscript.

\end{document}